\documentclass[12pt]{article}
\usepackage{geometry}
\usepackage{setspace}			
\usepackage{authblk}			
\usepackage{appendix}			
\usepackage{natbib}			
\usepackage[hidelinks]{hyperref}	

\usepackage{amsmath,amsfonts,amssymb}	
\usepackage{bm}					
\usepackage{bbm}			
\usepackage[bb=stix]{mathalpha}	
\usepackage{xfrac}				
\usepackage{mathtools}			

\usepackage{graphicx}			
\usepackage{caption}			
\usepackage{subcaption}			
\graphicspath{{./PNG files}}	
\usepackage{float}				
\usepackage[section]{placeins}	

\usepackage{longtable}			
\usepackage{booktabs}			
\usepackage{multirow}			
\usepackage{multicol}			

\usepackage{color}				
\usepackage{ulem}				
\usepackage{cancel}				

\newcommand{\tESG}{\text{ESG}}
\newcommand{\tX}{\text{X}}
\newcommand{\tM}{\text{M}}
\newcommand{\tu}{\text{u}}
\newcommand{\td}{\text{d}}

\begin{document}

\title{Path--dependent, ESG--valued, option pricing in the
Bachelier--Black--Scholes--Merton model}

\author{Bhathiya Divelgama\thanks{Corresponding author, bdivelga@ttu.edu}}
\author{Nancy Asare Nyarko}
\author{W. Brent Lindquist}
\author{Svetlozar T. Rachev}
\author{Blessing Omotade}
\affil{Department of Mathematics \& Statistics, Texas Tech University, Lubbock, TX, USA}

\date{}
\maketitle
\begin{abstract}
We extend the application of the Cherny--Shiryaev--Yor invariance principle
to a unified Bachelier--Black--Scholes--Merton (BBSM) dynamic pricing model.
This extension incorporates the influence of the history of the dynamics
(i.e., the path dynamics) of a market index on stock price changes.
We add an ESG rating component to the price of the risky asset (stock),
in such a manner that the impact of the ESG rating on the stock valuation
can be explored through variation in the value of a single parameter.
We develop discrete, binary tree, option pricing under this extended model.
Using an empirical data set of 10 stocks chosen from the Nasdaq-100,
we fit the model to stock price changes and compare model--based and published
European call option prices.

\end{abstract}

\section{Introduction}\label{sec:intro}
\cite{Hu_2020b} introduced a binomial, ``path dependent'', option pricing model where
the price of the underlying asset (stock) depends on the history of the dynamics
of a market influencing factor.
The intent of the model is to capture a more realistic interaction between an individual
asset and systemic market movement.
They assumed a Black--Scholes--Merton (BSM) model \citep{Black_1973, Merton_1973},
with the log-return of the stock price
driven by processes derived from application of the Cherny--Shiryaev--Yor invariance
principle \citep[CSYIP:][]{Cherny_2003} to the log--return of the market influencing
factor.
Under the assumption of time--independent coefficients,
use of the CSYIP enabled the identification of the continuum process driving the
stock price to which the discrete process converged weakly.

We consider this path dependent formalism under several extensions.
We extend the application of CSYIP to a unified Bachelier--Black--Scholes--Merton (BBSM)
pricing model.
This enables us to examine the effectiveness of a richer model which contains both the
classical BSM model and a Bachelier model as separate limits.
We also add an ESG rating component to the stock price, in such a manner that the impact
of the ESG rating on the stock price can be explored through variation in the value of a
single parameter.
Finally we consider application of the model to the pricing of European call options for
10 stocks chosen from the Nasdaq-100.
This makes the Nasdaq-100 index an obvious choice for the market factor influencing
the price of each stock.

\cite{lindquist2024} developed the unified BBSM
model, which incorporates both arithmetic and geometric Brownian dynamics into a
single coherent framework.
This unified BBSM model is market--complete, 
admits a unique risk--neutral measure and precludes arbitrage.
Unlike the BSM model, however, 
the BBSM model does not require asset prices or interest rates to remain non--negative.
Price change, rather than price return, is the fundamental ``change'' variable
of the BBSM model.
The standard BSM model is realized as the zero--value of a subset of the model
parameters.
Similarly, a modernized Bachelier's (MB) model is realized as the zero--value of a
(disjoint) subset of the model parameters.
\cite{lindquist2024} referred to this Bachelier model as modernized
as it corrected for deficiencies in past Bachelier models
\citep{Bachelier_1900,Shiryaev1999,schachermayer2008,Choi_2022,Brooks2017}.

As ESG ratings gained prominence, 
they transformed the landscape of financial markets and investment strategies. 
ESG factors, upon which the ratings are based,
reflect the sustainable and ethical practices of a firm, 
and have become a driving force behind (some) investor preferences, 
influencing company valuation and risk assessment.
For previous work incorporating ESG ratings into asset pricing see:
\cite{hu2024,lauria2022,Pastor2021,Pedersen2021,Rachev_2024,Zerbib2022,asare-nyarko2023};
and \cite{asare-nyarko2025}.
\cite{Rachev_2024} introduced an ESG affinity parameter $\gamma^{\tESG}$ into
the computation of an ESG--adjusted (ESG--valued) stock price.
This parameter controlled the impact of a company's ESG rating on the ESG--valued price
of its stock, with $\gamma^{\tESG} = 0$ indicating no influence;
a value of $\gamma^{\tESG} < 0$ indicating that the ESG rating was depressing the
stock's ESG-valuation; and $\gamma^{\tESG} > 0$ indicating the logical opposite.
The dynamics of the ESG--adjusted stock price was assumed to follow the MB model.
In their work, the model was investigated under both discrete--time and continuous--time
settings.
Subsequent research by \cite{asare-nyarko2025} embedded ESG--adjusted pricing into the
BBSM framework.
Focusing on empirical application, their study incorporated the binary option pricing
structure developed by \cite{lindquist2024}.
Basing their research on Nasdaq-100 stocks,
they provided empirical examples of ESG--adjusted prices and computed call option prices,
extracting implied ESG values to gauge sustainability preferences of option traders.

We develop the theoretical model in Section~\ref{sec:the_mdl}.
We briefly review the BBSM model of \cite{lindquist2024} in Section~\ref{sec:BBSM},
and the ESG--valued pricing model of \cite{Rachev_2024} in Section~\ref{sec:ESG}.
In Section~\ref{sec:PD_BBSM} we present our binary option pricing model under which
the price dynamics of an individual stock are influenced by the trajectory of a broader
market index.
In Section~\ref{sec:conv} we consider the weak convergence limits of the discrete model.
Section~\ref{sec:Emp} presents empirical results of the application of the option pricing
model to selected stocks from the Nasdaq-100.
Estimation of the market index driven processes derived from the CSYIP are presented
in Section~\ref{sec:CSYIP}.
The procedure for, and results of, fitting the parameters of the path--dependent BBSM
model are presented in Section~\ref{sec:params}.
The results for the model computation of European call option prices for the selected
stocks are presented in Section~\ref{sec:op_prc}.
Final conclusions appear in Section~\ref{sec:conc}.

\section{The Model}\label{sec:the_mdl}

\subsection{Review of the BBSM model}\label{sec:BBSM}
The \cite{lindquist2024} BBSM market model $(\mathcal{A,B,C})$ consists, as usual, of a risky asset 
$\mathcal{A}$, a riskless asset $\mathcal{B}$, 
and a European contingent claim (option) $\mathcal{C}$. 
The risky asset $\mathcal{A}$ has the price dynamics of a continuous
diffusion process determined by the stochastic differential equation
\begin{subequations}
\begin{align}
	dA_t &= \varphi_t dt + \psi_t dB_t \, , \qquad t \geq 0 \, ,   \quad A_0 > 0, \label{eq1.3a}\\
	\varphi_t &= a_t + \mu_t A_t, \qquad  \psi_t = v_t + \sigma_t A_t, \label{eq1.3b}
\end{align}
\end{subequations}
where $B_t$, $t \in [0, \infty)$, 
is a standard Brownian motion on a stochastic basis (filtered probability space) 
$(\Omega,\mathcal{F}, \mathbb{F} = \{\mathcal{F}_t = \sigma(B_u, u \leq t) \subseteq F, t \geq 0\},
\mathbb{P})$ 
on a complete probability space $(\Omega, \mathcal{F}, \mathbb{P})$.
The processes 
$a_t \in \mathbb{R}$, $\mu_t \in \mathbb{R}$, $v_t \in \mathbb{R}$, and $\sigma_t \in \mathbb{R}$,
are $\mathcal{F}$--adapted for $t \geq 0$ and satisfy the usual regularity conditions.\footnote{
	See \citet[Section 5G and Appendix E]{duffie2001}. 
	The regularity conditions are satisfied if:
	$\varphi_t$ and $\psi_t$ are $\mathbb{F}$--adapted for $t\geq 0$; 
	satisfy Lipschitz and growth conditions for $x \in \mathbb{R}$; and 
	$\int_{0}^{t} |\varphi_t| \, ds < \infty$, $\int_{0}^{t} |\psi_t^2| \, ds < \infty$
	for all $t \geq 0$. 
	To simplify the exposition, we will assume that $a_t$, $\mu_t$, $v_t$, and $\sigma_t$
	have trajectories that are continuous and uniformly bounded on $[0,\infty)$.
}
The parameters $a_t$, $\mu_t$, $v_t$ and $\sigma_t$ define a four--dimensional (4D)
space.
In the two--dimensional (2D) subspace, $a_t = v_t = 0$, \eqref{eq1.3a} reduces
to geometric Brownian motion with variable coefficients,
which would require the volatility satisfy $\sigma_t \geq 0$, $\mathbb{P}$--almost
surely.
In the (disjoint) 2D subspace  $\mu_t = \sigma_t = 0$, \eqref{eq1.3a} reduces to
arithmetic Brownian motion, which would require
the volatility satisfy $v_t \geq 0$, $\mathbb{P}$--almost surely.
For the model \eqref{eq1.3a}, the volatility must satisfy
\begin{equation}\label{eq:BBSM_vol}
	\psi_t = v_t + \sigma_t A_t \geq 0, \quad \mathbb{P}\text{--almost surely}.
\end{equation}
We note that imposition of \eqref{eq:BBSM_vol} also imposes the required restrictions on
$v_t$ and $\sigma_t$ in the appropriate subspaces.\footnote{
	Since geometric Brownian motion only develops positive prices, then the requirement
	$v_t + \sigma_t A_t \geq 0 \xrightarrow[v_t = 0]{} \sigma_t \geq 0$.
}
Therefore the BBSM model only requires the risky--asset volatility restriction
\eqref{eq:BBSM_vol}.

Consistent with \eqref{eq1.3a},\footnote{
	As argued in \cite{lindquist2024}, \eqref{eq1.4} should have the form \eqref{eq1.3a}
	with zero volatility.
	We note additionally that any other definition of \eqref{eq1.4} would effectively
	lead to two distinct riskless assets, resulting in arbitrage opportunities.
}
the riskless asset $\mathcal{B}$ has the dynamics
\begin{equation}\label{eq1.4}
	d\beta_t = \chi_t dt, \qquad \chi_t = \rho_t + r_t \beta_t,  \quad t \geq 0, \quad \beta_0 > 0,
\end{equation}
where $\rho_t \in \mathbb{R}$ and $r_t \in \mathbb{R}$.
The full BBSM model has a 6D parameter space determined by $a_t$, $\mu_t$, $v_t$,
$\sigma_t$, $\rho_t$, and $r_t$.
The 3D subspace $\mu_t = \sigma_t = r_t = 0$ reduces the BBSM model to the MB model,
whereas the 3D subspace $a_t = v_t = \rho_t = 0$ reduces to the usual BSM model.
The no--arbitrage assumption leads to the market price of risk
$\Theta_t = (\varphi_t - \chi_t) / \psi_t$
being strictly positive for all $t \geq 0$.
As $\psi_t$ is required to be positive $\mathbb{P}$--almost surely,
then $\varphi_t - \chi_t$ must also be positive $\mathbb{P}$--almost surely.

The solution of the linear stochastic differential equation defined by
\eqref{eq1.3a} and \eqref{eq1.3b} is
\begin{equation}\label{eq1.5}
  \begin{aligned}
	A_t &= \xi(t) \left[A_0 + \int_0^t \frac{1}{\xi(u)} (a_u - v_u \sigma_u) du
 				+ \int_0^t \frac{v_u}{\xi(u)} dB_u \right], \quad t \geq 0,\\
	&\qquad \xi(t) = \exp \left\{\int_0^t \left( \mu_u - \frac{\sigma_u^2}{2} \right) du
						+ \int_0^t \sigma_u dB_u\right\} .
  \end{aligned}
\end{equation}
The solution to \eqref{eq1.4} is 
\begin{equation}\label{eq1.6}
	\beta_t = \beta_0 e^{\int_0^t r_u du} + \int_0^t e^{\int_u^t 
	r_s ds} \rho_u du.
\end{equation}
Examination of the solutions \eqref{eq1.5} and \eqref{eq1.6} reveals, 
depending on parameter values,
the possibility of negative prices.\footnote{
	For example, $\rho_t < 0$ can lead to negative values for $\beta_t$.
}

The option $\mathcal{C}$ has the price $C_t = f\left( A_t,t \right)$,
where $f(x,t)$, $x \in \mathbb{R}$, $t \in [0,T]$
has continuous partial derivatives $\partial^{2}f(x,t) / \partial x^2$
and $\partial f(x,t) / \partial t$ on $t \in [0,T]$;
$T$ is the terminal (maturity) time of $\mathcal{C}$;
and the option’s terminal payoff is $C_T = g\left( A_T \right)$
for some continuous function $g:\mathbb{R}\rightarrow\mathbb{R}$.
From It{\^o}’s formula,
\begin{equation}\label{eq:C_Ito}
	df\left( A_t,t \right) = \left[
		\frac{\partial f\left( A_t,t \right)}{\partial t}
		+ \phi_t \frac{\partial f\left( A_t,t \right)}{\partial x}
		+ \frac{\psi_t^2}{2} \frac{\partial^{2}f\left( A_t,t \right)}{\partial x^2}
		\right] dt
		+ \frac{\partial f\left( A_t,t \right)}{\partial x}\psi_t dB_t.
\end{equation}
Using the classical treatment of a replicating, self--financing portfolio,
\cite{lindquist2024} developed the risk-neutral partial differential equation (PDE)
for $f(\cdot)$ in the cases in which the riskless asset is a) a bond or
b) a bank account, and c) when the risky asset pays a dividend.
In each case they analyzed the Feynman--Kac solution to this PDE for the fair price
of the option.

\subsection{ESG--Adjusted Pricing}\label{sec:ESG}
The \cite{Rachev_2024} ESG-adjusted price for the stock of company X is
\begin{equation}\label{eq:ESG_prc}
	A_t^{(\tX)} = S_t^{(\tX)} \left( 1 + \gamma^{\tESG} Z_t^{(\tX;\tM)} \right),
\end{equation}
where $S_t^{(\tX)}$ is the quoted financial stock price and $\gamma^{\tESG} \in \mathbb{R}$
is the ESG affinity of the financial market, quantifying the market view of the impact
of ESG ratings on asset values.
The final term in \eqref{eq:ESG_prc} is the relative ESG rating
\begin{equation}\label{eq:ESG_rel}
	Z_t^{(\tX;\tM)} = \frac{Z_t^{(\tX)} - Z_t^{(\tM)}}{Z_t^{(\tM)}},
\end{equation}
where $Z_t^{(\tX)}$ is the ESG rating of the company and $Z_t^{(\tM)}$ is
the ESG rating of the market index.\footnote{
	Typically $Z_t^{(\tX)}$ is quoted on a zero---to--ten or a zero--to--one--hundred scale.
	Our provider, Bloomberg Professional Services, uses the zero--to--ten scale.
	Using \eqref{eq:ESG_rel}, the value of $Z_t^{(\tX;\tM)}$ is independent of the scale,
	as long as the range of this scale is finite and the same range is used for $Z_t^{(\tX)}$
	and $Z_t^{(\tM)}$.
	The ESG rating $Z_t^{(\tM)}$ for the index can be computed using an appropriately
	weighted average of the ESG ratings of the stocks comprising the index.
}
Equation \eqref{eq:ESG_prc} determines the following dependencies of $A_t^{(\tX)}$ on
$\gamma^{\tESG}$:
\begin{equation}\label{eq:ESG_dep}
  \begin{aligned}
	\mathbb{E}\left[ A_t^{(\tX)} \right] &= \mathbb{E}\left[ S_t^{(\tX)} \right]
		+ \gamma^{\tESG} \mathbb{E} \left[ S_t^{(\tX)} Z_t^{(\tX;\tM)} \right];\\
	\text{Var}\left[ A_t^{(\tX)} \right] &= \text{Var}\left[ S_t^{(\tX)} \right]
		+ 2 \gamma^{\tESG} \text{Cov} \left[ S_t^{(\tX)}, S_t^{(\tX)} Z_t^{(\tX;\tM)} \right]
		+ \left( \gamma^{\tESG} \right)^2 \text{Var}\left[ S_t^{(\tX)} Z_t^{(\tX;\tM)}  \right];\\
	A_{t+1}^{(\tX)} - A_t^{(\tX)} &= S_{t+1}^{(\tX)} - S_t^{(\tX)} 
		+ \gamma^{\tESG} \left( S_{t+1}^{(\tX)} Z_{t+1}^{(\tX;\tM)} - S_t^{(\tX)} Z_t^{(\tX;\tM)} \right).
  \end{aligned}
\end{equation}

The model \eqref{eq:ESG_prc} has been subjected to several criticisms.
The first is that the ESG valuation $A_t^{(\tX)}$ is not a financial price;
the stock is bought and sold under the price $S_t^{(\tX)}$.
The response to this criticism is that $A_t^{(\tX)}$ need not be a price for it to be
a useful valuation under which to determine, for example, portfolio selection.
Indeed, \cite{lauria2022} have developed portfolio optimization under BSM using a similar
ESG--adjusted pricing model, while \cite{hu2024}, \cite{asare-nyarko2023}
and \cite{asare-nyarko2025}
have developed option pricing models for ESG--valued hedging using \eqref{eq:ESG_prc}.
As $A_t^{(\tX)} = S_t^{(\tX)}$ when $\gamma^{\tESG} = 0$, for brevity we will refer to
$A_t^{(\tX)}$ as a ``price''.

A second criticism is that the ESG ratings, and their component risks,
are already reflected in the market--determined price of $S_t^{(\tX)}$.
The response to this criticism is ``clearly no'', given the anti--ESG,
anti--climate--change views pervading current US finance industry and political
views.
The pricing of stocks of heavy polluters such as the oil and gas industry and of
massive electricity users such as the information technology sector
is not commensurate with their climate change risk.
Political reduction of protections imposed on large financial banks will increase
governance risk, while reductions on EPA and CDC protections increase environmental
and social risk.
Essentially all of these factors are being ignored by current US markets in order
to focus on increased financial profit.

As any specific stock discussed will be clearly identified from the context of the
discussion, we will drop the company designation X from the ESG--adjusted price and simply refer to
$A_t$.

\subsection{BBSM Path--Dependent Model}\label{sec:PD_BBSM}
Employing the CSYIP, we develop our binary, path--dependent, BBSM option pricing model
for a stock whose dynamics are influenced by the trajectory of a market index.
Our discrete market model is extended to be $(\mathcal{A,B,C,M})$,
where $\mathcal{M}$ is a market index.\footnote{
	In our empirical computations we utilized an index-tracking ETF rather than
	the index itself.
}
Both $\mathcal{A}$ and $\mathcal{M}$ will be driven by the same Brownian motion,
so the extended market model is complete.
We again note that price dynamics in the BBSM model are developed for price changes
and not for returns.

Let
$A_{t_{n,k}}^{(\tM)}$, $t_{n,k} = k\Delta$, $k = 1, \ldots, n$, 
$n\Delta = T$, $n \in \mathbb{N}$, $A_{t_{n,0}}^{(\tM)} = A_0^{(\tM)}$, 
be the value of $\mathcal{M}$ over the period $[t_{n,k}, t_{n,k+1})$.  
Let 
$c_{t_{n,k}}^{(\tM)} = A_{t_{n,k}}^{(\tM)} - A_{t_{n,k-1}}^{(\tM)}$
denote the change in the value of $\mathcal{M}$ over the $k$'th time period
$[t_{n,k-1}, t_{n,k}]$,
and $\varsigma_{n,k}^{(\tM)}$, $k = 1, \ldots, n$, 
denote the binary sequence of the directions of these changes over time,
\begin{equation*}
	\varsigma_{n,k}^{(\tM)} = 
	\begin{cases} 
		\ \ \,1, & \text{if } c_{t_{n,k}}^{(\tM)} \geq 0, \\
		-1, & \text{if } c_{t_{n,k}}^{(\tM)} < 0.
	\end{cases}
\end{equation*}
We assume that the $\varsigma_{n,k}^{(\tM)}$, $k = 1, \ldots, n,$ 
are independent random signs with 
\begin{equation*}
	\mathbb{P}\left(\varsigma_{n,k}^{(\tM)} = 1 \right) = 
	1 - \mathbb{P}\left(\varsigma_{n,k}^{(\tM)} = -1 \right)
		= p_{n,\Delta}^{(\tM)}
		= p_0^{(\tM)} + p_1^{(\tM)} \sqrt{\Delta} + p_2^{(\tM)}\Delta, 
\end{equation*}
where 
$p_0^{(\tM)} \in (0,1)$, $p_1^{(\tM)} \in \mathbb{R}$, $p_2^{(\tM)} \in \mathbb{R}$.
This ensures that $p_{n,\Delta}^{(\tM)} \in (0,1)$ as $\Delta = \frac{T}{n} \downarrow 0$.
(In empirical applications, $p_{n,\Delta}^{(\tM)}$ will be computed from a real market index,
which will satisfy $p_{n,\Delta}^{(\tM)} \in (0,1)$.
The Bernoulli variables $\varsigma_{n,k}^{(\tM)}$ define a filtration,
$$
	\mathbb{F}^{(n,\tM)} = \left\{ \mathcal{F}^{(n,\tM)}_k  = \sigma \left( \varsigma_{n,j}^{(\tM)}, \ j = 1, ..., k \right),
	\ \ k = 1, ..., n,\ \mathcal{F}^{(n,\tM)}_0 = \{\emptyset,\Omega\},\ \varsigma_{n,0}^{(\tM)} = 0 \right\},
$$
and the stochastic basis
$\left\{ \Omega^{(\tM)}, \mathcal{F}^{(\tM)}, \mathbb{F}^{(n,\tM)}, \mathbb{P}\right\}$
on the complete probability space $\big\{ \Omega^{(\tM)}, \mathcal{F}^{(\tM)}, \mathbb{P} \big\}$.

We model the discrete dynamics of the market index
following the BBSM formulation outlined in \citet[][Equations (34)–(36)]{lindquist2024}:
\begin{equation}\label{eq4.4}
	\begin{aligned}
		A_{t_{n,k+1}}^{(\tM)} 
		&= A_{t_{n,k}}^{(\tM)} + c_{t_{n,k+1}}^{(\tM)},
			\quad k=0,\ldots, n-1,\\
		A_{t_{n,0}}^{(\tM)} &= A_0^{(\tM)} > 0,
	\end{aligned}
\end{equation}
where 
\begin{equation}\label{eq4.5}
   c_{t_{n,k+1}}^{(\tM)} = 
  \begin{cases} 
	c_{t_{n,k+1}}^{(M,u)} = \varphi_{t_{n,k}}^{(\tM)} \Delta
		+ \sqrt{\frac{1 - p_{n,\Delta}^{(\tM)}}{\strut p_{n,\Delta}^{(\tM)}}}\  \psi_{t_{n,k}}^{(\tM)} \sqrt{\Delta}, 
		& \text{if } \varsigma_{n,k}^{(\tM)} = 1, \\
	c_{t_{n,k+1}}^{(M,d)} = \varphi_{t_{n,k}}^{(\tM)} \Delta
		- \sqrt{\frac{p_{n,\Delta}^{(\tM)}}{\strut 1 - p_{n,\Delta}^{(\tM)}}}\  \psi_{t_{n,k}}^{(\tM)} \sqrt{\Delta}, 
		& \text{if } \varsigma_{n,k}^{(\tM)} = 0,
  \end{cases} 
\end{equation}
with
\begin{equation}\label{eq4.6}
	\varphi_{t_{n,k}}^{(\tM)} = a_{t_{n,k}}^{(\tM)} +    \mu_{t_{n,k}}^{(\tM)} A_{t_{n,k}}^{(\tM)}, \qquad
	   \psi_{t_{n,k}}^{(\tM)} = v_{t_{n,k}}^{(\tM)} + \sigma_{t_{n,k}}^{(\tM)} A_{t_{n,k}}^{(\tM)}.
\end{equation}
A careful analysis shows that a price process of the form \eqref{eq4.5}, \eqref{eq4.6}
is not recombining \citep{asare-nyarko2025}; it develops as a binary tree.

We apply the CSYIP framework (this framework is briefly summarized in Appendix~\ref{app:CSYIP}),
defining
\begin{equation}\label{eq4.7}
  \begin{aligned}
	&\text{(i)} \quad Z_{t_{n,k}}^{(\tM)}
		= \frac{1}{\sqrt{\text{Var}\left[c_{t_{n,k}}^{(\tM)}\right]}}
			\left(c_{t_{n,k}}^{(\tM)} - \mathbb{E}\left[c_{t_{n,k}}^{(\tM)}\right]\right),
			\quad k = 1, ..., n,\\
	&\text{(ii)} \quad \xi_{n,k}^{(\tM)}
		= \sqrt{\frac{1 - \mathbb{p}_{n,\Delta}^{(\tM)}}{\mathbb{p}_{n,\Delta}^{(\tM)}}} \, 
			\text{I}_{ \left\{Z_{t_{n,k}}^{(\tM)} \geq 0 \right\} } 
			- \sqrt{\frac{\mathbb{p}_{n,\Delta}^{(\tM)}}{1 - \mathbb{p}_{n,\Delta}^{(\tM)}}} \,
			\text{I}_{ \left\{Z_{t_{n,k}}^{(\tM)} < 0 \right\} }, \quad k = 1, ..., n,\\
	&\qquad \mathbb{p}_{n,\Delta}^{(\tM)}
		= \mathbb{P}\left(Z_{t_{n,k}}^{(\tM)} \geq 0 \right) \in (0,1), \\
	&\text{(iii)} \quad X_{k/n}^{(\tM)} = \sum_{i=1}^{k} \sqrt{\Delta} \xi_{n,i}^{(\tM)},
		\quad k = 1, ..., n,\qquad X_0^{(\tM)} = 0,\\
	&\text{(iv)} \quad Y_{k/n}^{(M,h)} = \sum_{i=1}^{k} \sqrt{\Delta} \xi_{n,i}^{(\tM)}
											h\left( X_{(i-1)/n}^{(\tM)} \right),
		\quad k = 1, ..., n, \qquad Y_0^{(M,h)} = 0.
  \end{aligned}   
\end{equation}
In (ii), $\textrm{I}_{ \{\cdot\} }$ is the indicator function;
in (iv), $h(x)$ is any CSY piecewise continuous function.
We note that,
while $p_{n,\Delta}^{(\tM)}$ is the probability for an upturn of the index
over each time interval $[t_{n,k-1}, t_{n,k}]$,
$\mathbb{p}_{n,\Delta}^{(\tM)}$ is the probability for an upturn of the
centralized difference
$c_{t_{n,k}}^{(\tM)} - \mathbb{E}\left[c_{t_{n,k}}^{(\tM)}\right]$ in the index value.
By (ii), the Bernoulli random variables $ \xi_{n,k}^{(\tM)}$ satisfy
$\mathbb{E}[\xi_{n,k}^{(\tM)}] = 0$ and $\text{Var}[\xi_{n,k}^{(\tM)}] = 1$,
and are further assumed to be independent and identically distributed (iid).
Thus (see Section~\ref{sec:conv}) $X_{k/n}^{(\tM)}$ is,
under the assumptions on $\xi_{n,k}^{(\tM)}$,
a discrete approximation to a Brownian motion trajectory having the increment
$\sqrt{\Delta}\, \xi_{n,i}^{(\tM)}$ over the time step $[t_{n,i-1}, t_{n,i}]$, $i = 1, ..., k$.

Consider the discrete filtration 
\begin{equation*}
	\mathbb{F}^{(d)} = \left( \mathcal{F}_0 = \{\emptyset, \Omega\}, 
		\ \mathcal{F}_k = \sigma(\xi_{n,1}^{(\tM)}, \ldots, \xi_{n,k}^{(\tM)}),
		\ k = 1, \ldots, n,\  n \in \mathbb{N} \right).
\end{equation*}
Conditional on $\mathcal{F}_k$,
our binary, path--dependent model for the price of the risky asset is
\begin{equation}\label{eq5.2}
	c_{t_{n,k+1}}
		= \varphi_{t_{n,k}} \Delta 
		+ \psi_{t_{n,k}}\, \sqrt{\Delta}\, \xi_{n,k+1}^{(\tM)} 
		+ \gamma_{t_{n,k}}\, \sqrt{\Delta}\, \xi_{n,k+1}^{(\tM)}\,
				h\left( X_{k/n}^{(\tM)} \right).
\end{equation}
Expanding $\varphi_{t_{n,k}}$ and $\psi_{t_{n,k}}$ using the discrete form of \eqref{eq1.3b},
\eqref{eq5.2} can be written
\begin{equation}\label{eq:c_tilde}
  \begin{aligned}
	c_{t_{n,k+1}} 
		&=  a_{t_{n,k}} \Delta + \mu_{t_{n,k}} \Delta A_{t_{n,k}}
			+  v_{t_{n,k}} \sqrt{\Delta}\,\xi_{n,k+1}^{(\tM)}
			+  \sigma_{t_{n,k}} \sqrt{\Delta}\, A_{t_{n,k}} \xi_{n,k+1}^{(\tM)}\\
		&\qquad + \gamma_{t_{n,k}}\, \sqrt{\Delta}\,
					h\left( X_{k/n}^{(\tM)}  \right)  \xi_{n,k+1}^{(\tM)}\\
		&= \tilde{a}_{t_{n,k}} + \tilde{\mu}_{t_{n,k}} A_{t_{n,k}} + \tilde{v}_{t_{n,k}} \xi_{n,k+1}^{(\tM)}
		+ \tilde{\sigma}_{t_{n,k}} A_{t_{n,k}} \xi_{n,k+1}^{(\tM)}
		+ \tilde{\gamma}_{t_{n,k}} h\left( X_{k/n}^{(\tM)} \right) \xi_{n,k+1}^{(\tM)},
  \end{aligned}
\end{equation}
where
\begin{equation*}
	\tilde{a}_{t_{n,k}} = a_{t_{n,k}} \Delta, \quad
	\tilde{\mu}_{t_{n,k}} = \mu_{t_{n,k}} \Delta,\quad
	\tilde{v}_{t_{n,k}} = v_{t_{n,k}}\sqrt{\Delta}, \quad
	\tilde{\sigma}_{t_{n,k}} = \sigma_{t_{n,k}}\sqrt{\Delta}, \quad
	\tilde{\gamma}_{t_{n,k}} = \gamma_{t_{n,k}} \sqrt{\Delta}\,.
\end{equation*}

The last equality in \eqref{eq:c_tilde} defines $c_{t_{n,k+1}}$ as a regression against the
independent terms $A_{t_{n,k}}$, $\xi_{n,k+1}^{(\tM)}$, $A_{t_{n,k}} \, \xi_{n,k+1}^{(\tM)}$
and $h\left( X_{k/n}^{(\tM)} \right) \xi_{n,k+1}^{(\tM)}$.
Equation \eqref{eq:c_tilde} (equivalently \eqref{eq5.2}) implies a dependence relationship
between $A_{t_{n,k}}$ and the form of the function $h\left( X_{k/n}^{(\tM)} \right)$.
However, this relationship is subtle.
From a pure model perspective, deciding on the market index $\tM$, determining a choice
for $h(\cdot)$,
and determining the coefficient set $\tilde{a}_{t_{n,k}}, ..., \tilde{\gamma}_{t_{n,k}}$
(most easily done assuming time--independent coefficients),
certainly defines a price change process $c_{t_{n,k}}$  (and trivially $A_{t_{n,k}}$).
Changing the form of $h(\cdot)$ will change the price process, implying a dependence of
$A_{t_{n,k}}$ on $h(\cdot)$.
However, from an empirical perspective, the time series $c_{t_{n,k}}$ and
$\xi_{n,k}^{(\tM)}$ are given (from the selected stock and market index),
and the problem is to determine an appropriate function
$h\left( X_{k/n}^{(\tM)} \right)$ and the coefficient set
$\tilde{a}_{t_{n,k}}, ..., \tilde{\gamma}_{t_{n,k}}$
that provide the best fit of the model to these time series.
Thus from an empirical perspective, it is the function $h(\cdot)$ that depends on the
choice of stock and market index.
The goal is to find a functional form that ``works well'' for a wide variety of stocks
for a given market index.
As we argue in Section~\ref{sec:CSYIP}, the form of $h(\cdot)$ is mostly dependent on
the index itself.

We consider the estimation of the coefficients in the regression problem \eqref{eq:c_tilde}
in Section~\ref{sec:params};
continuing with form \eqref{eq5.2} is more convenient for the discussion of option pricing.
From the CSYIP property (ii), we define
\begin{equation*}
	\xi_\tu = \sqrt{\frac{1 - \mathbb{p}_{n,\Delta}^{(\tM)}}{\mathbb{p}_{n,\Delta}^{(\tM)}}}\,, \qquad
	\xi_\td = \sqrt{\frac{\mathbb{p}_{n,\Delta}^{(\tM)}}{1 - \mathbb{p}_{n,\Delta}^{(\tM)}}}.
\end{equation*}
Thus the binary model \eqref{eq5.2} can be written
\begin{equation}\label{eq:BBSM_bin}
	c_{t_{n,k+1}} = 
	\begin{cases}
		c_{t_{n,k+1}}^{(\tu)} = \varphi_{t_{n,k}} \Delta + \left[ \psi_{t_{n,k}} + \gamma_{t_{n,k}}
			h\left( X_{k/n}^{(\tM)} \right) \right]
			\sqrt{\Delta}\, \xi_\tu, \text{ w.p. } \mathbb{p}_{n,\Delta}^{(\tM)},\\[10pt]
		c_{t_{n,k+1}}^{(\td)} = \varphi_{t_{n,k}} \Delta - \left[ \psi_{t_{n,k}} + \gamma_{t_{n,k}}
			h\left( X_{k/n}^{(\tM)} \right) \right]
			\sqrt{\Delta}\, \xi_\td, \text{ w.p. } 1 - \mathbb{p}_{n,\Delta}^{(\tM)}.
	\end{cases}
\end{equation}
Conditional on $\mathcal{F}_k$, the mean and variance of $c_{t_{n,k+1}}$ are
\begin{equation}\label{eq:c_mean_var}
  \begin{aligned}
	\mathbb{E}\left[ c_{t_{n,k+1}} \left | \mathcal{F}_k \right. \right]
		&= \varphi_{t_{n,k}} \Delta,\\
	\text{Var}\left[ c_{t_{n,k+1}} \left | \mathcal{F}_k \right. \right]
		&= \left\{ \psi_{t_{n,k}} + \gamma_{t_{n,k}} h\left( X_{k/n}^{(\tM)} \right) \right\}^2 \Delta.
  \end{aligned}
\end{equation}
For brevity, we denote the conditional volatility of $c_{t_{n,k+1}}$ as
$$
	\eta_{k,\Delta}^{(\tM)} = \psi_{t_{n,k}} + \gamma_{t_{n,k}} h\left( X_{k/n}^{(\tM)} \right).
$$
In contrast to \eqref{eq:BBSM_vol},
the volatility requirement for the path--dependent model is
\begin{equation}\label{eq:PD_vol}
	\eta_{k,\Delta}^{(\tM)} \geq 0, \quad \mathbb{P}\text{--almost surely}.
\end{equation}
The path dependence of the market index impacts the conditional volatility.
If $ \gamma \neq 0 $, 
the risky asset price change $c_{t_{n,k+1}}$
depends on the entire sequence (path) $\xi_{n,i}^{(\tM)}$, $i=1, ... ,k$,
of up-- and downturns of the market index prior to $t_{n,k+1}$.
Consequently, the price of the risky asset will not be a Markov process.

The riskless asset $\mathcal{B}$ has the discrete price dynamics
\begin{equation}\label{eq:B_dyn}
	\beta_{t_{n,k+1}} = \beta_{t_{n,k}} + \chi_{t_{n,k}} \Delta
		=  \tilde{\rho}_{t_{n,k}} + (1 + \tilde{r}_{t_{n,k}}) \beta_{t_{n,k}},
		\qquad k = 0, ..., n-1, \quad \beta_0 > 0,
\end{equation}
where $\tilde{\rho}_{t_{n,k}} = \rho_{t_{n,k}} \Delta$ and
$\tilde{r}_{t_{n,k}} = r_{t_{n,k}} \Delta$.\footnote{
	Equation \eqref{eq:B_dyn} has been written both in a form useful for
	theoretical discussion as well as for regression fitting.
}

The option $\mathcal{C}$ has the discrete price dynamics
$C_{t_{n,k}} = f(A_{t_{n,k}}, k\Delta)$, $k = 0, ..., n-1$,
with terminal value $C_{t_{n,n}} = g(A_{t_{n,n}})$.
Consider a self--financing strategy,
$P_{t_{n,k}} = \mathbbm{a}_{t_{n,k}} A_{t_{n,k}} + \mathbbm{b}_{t_{n,k}}\beta_{t_{n,k}}$,
replicating the option price $C_{t_{n,k}}$.
Solving the replicating system of equations in the usual manner produces the recursion
formula for the risk neutral valuation of the option price on the binary tree,
\begin{equation}\label{eq5.5}
 	C_{t_{n,k}} = \frac{\beta_{t_{n,k}}} {\beta_{t_{n,k+1}}}
		\left[\mathbb{q}_{t_{n,k}} C_{t_{n,k+1}}^{(\tu)}
			+ (1 - \mathbb{q}_{t_{n,k}}) C_{t_{n,k+1}}^{(\td)}\right],
\end{equation}
where $C_{t_{n,k+1}}^{(\tu)}$ and $C_{t_{n,k+1}}^{(\td)}$ represent the ``up''
and ``down'' values on the binary tree contributing to the option value $C_{t_{n,k}}$
at a given node on the tree.
The conditional risk--neutral probability is given by 
\begin{equation}\label{eq5.3}
  \begin{aligned}
	\mathbb{q}_{t_{n,k}} &=
	\frac{
	\left[ \eta_{k,\Delta}^{(\tM)} \xi_{\td}
		+ \left( \cfrac{\chi_{t_{n,k}}}{\beta_{t_{n,k}}} A_{t_{n,k}} - \varphi_{t_{n,k}} \right) \sqrt{\Delta}
	\right] \sqrt{ \mathbb{p}_{n,\Delta}^{(\tM)} ( 1 - \mathbb{p}_{n,\Delta}^{(\tM)} ) }
	} { \eta_{k,\Delta}^{(\tM)} }\\
	&= \mathbb{p}_{n,\Delta}^{(\tM)}
		- \frac{ \left( \varphi_{t_{n,k}} - A_{t_{n,k}} \cfrac{\chi_{t_{n,k}}} {\beta_{t_{n,k}}} \right) }
			   { \eta_{k,\Delta}^{(\tM)} }
		\sqrt{ \mathbb{p}_{n,\Delta}^{(\tM)} ( 1 - \mathbb{p}_{n,\Delta}^{(\tM)} ) \Delta}\,.
  \end{aligned}
\end{equation}

\subsection{Convergence Limits of the Model}\label{sec:conv}
Set
\begin{equation*}
  \begin{aligned}
	\mathbb{A}_{[0,T]}^{(n,\tM)} &= \left\{ A_t^{(n,\tM)} = A_{t_{n,k}}^{(\tM)},
		\quad t \in [t_{n,k}, t_{n,k+1}), \quad k = 0, \ldots, n-1,\  
								A_T^{(n,\tM)} = A_{t_{n,n}}^{(\tM)} \right\},\\
	\mathbb{A}_{[0,T]}^{(\tM)} &= \{ A_t^{(\tM)}, \quad t \in [0,T] \},
  \end{aligned}
\end{equation*}
where $A_t^{(\tM)}$ is governed by the same dynamics \eqref{eq1.3a} as the risky asset
while $A_{t_{n,k}}^{(\tM)}$ is governed by \eqref{eq4.4}.
Then $\mathbb{A}_{[0,T]}^{(n,\tM)}$ converges weakly in 
$\left(\mathcal{D}[0,T], d^{(0)}\right)$ to $\mathbb{A}_{[0,T]}^{(\tM)}$.\footnote{
	The arguments for weak convergence are similar to those in
	\cite{Davydov2008} and \cite{Hu_2020}.
}

Consider the following processes in the Skorokhod space $ \mathcal{D}[0,T]$,
\begin{equation}\label{eq4.8}
\begin{aligned}
	\mathbb{B}_{[0,T]}^{(n)} &= \left\{B_t^{(n)} \ \ = X_{k/n}^{(\tM)},
		\quad \ t \in [t_{n,k}, t_{n,k+1}),\  B_T^{(n)} \ \ = X_1^{(\tM)} \right\}, \\
	\mathbb{C}_{[0,T]}^{(n,h)} &= \left\{C_t^{(n,h)} = Y_{k/n}^{(M,h)},
		\quad t \in [t_{n,k}, t_{n,k+1}),\  C_T^{(n,h)} = Y_1^{(M,h)} \right\}.
\end{aligned}
\end{equation}
Under the assumptions that the $\xi_{n,k}^{(\tM)}$ in the CSYIP property (ii) are iid,
as $n \to \infty$ the CSYIP states that the
bivariate process $\left( \mathbb{B}_{[0,T]}^{(n)}, \mathbb{C}_{[0,T]}^{(n,h)} \right)$ 
converges weakly in $\mathcal{D}[0,T] \times \mathcal{D}[0,T]$
to $\left( \mathbb{B}_{[0,T]}, \mathbb{C}_{[0,T]}^{(h)} \right)$,
where
$\mathbb{B}_{[0,T]} = \{B_t,\  t \in [0,T]\}$ is a Brownian motion on $[0,T]$,
and
$\mathbb{C}_{[0,T]}^{(h)} = \left\{C_t^{(h)} = \int_0^t h(B_s) \, dB_s,\  t \in [0,T] \right\}$.
Therefore, set
\begin{equation}\label{eq:PD_D0T}
  \begin{aligned}
	\mathbb{A}_{[0,T]}^{(n,h)} &= \left\{ A_t^{(n,h)} = A_{t_{n,k}},
		\quad t \in [t_{n,k}, t_{n,k+1}), \quad k = 0, \ldots, n-1,\  
								A_T^{(n} = A_{t_{n,n}} \right\},\\
	\mathbb{A}_{[0,T]}^{(h)} &= \left\{ A_t^{(h)}, \quad t \in [0,T] \right\},
  \end{aligned}
\end{equation}
where $A_t^{(h)}$ is governed by the dynamics
\begin{equation}\label{eq:Ah_dyn}
	dA_t^{(h)} = \varphi_t dt + \psi_t dB_t + \gamma_t h(B_t) \, dB_t,
	\qquad t \in [0,T],   \quad A_0^{(h)} > 0, 
\end{equation}
and $A_{t_{n,k}}$ is governed by \eqref{eq5.2}.
Then $\mathbb{A}_{[0,T]}^{(n,h)}$ converges weakly in $\left(\mathcal{D}[0,T]\right)$
to $\mathbb{A}_{[0,T]}^{(h)}$.

We note that \eqref{eq:Ah_dyn} falls outside of the class of Markovian It{\^o} stochastic
differential equations covered by \citet[Section 5G and Appendix E]{duffie2001}.
We leave as an open question the regularity conditions required for the existence of a
solution to \eqref{eq:Ah_dyn}.

\section{Empirical Results}\label{sec:Emp}
We considered a data set of 10 stocks chosen from Nasdaq-100 companies.
The stocks are listed in Appendix Table~\ref{tab:stocks}.
Price data spanned from 04 January 2016 through 02 January 2024.\footnote{
	Source: Bloomberg Professional Services. Accessed 01 March 2024.
}
At the date of access, fiscal year (FY) ESG scores were available for 2015
through 2022. 
Therefore, we conservatively assumed that each firm’s FY 2023 ESG score
remained unchanged from its FY 2022 value.
Appendix Table~\ref{tab:ESG} presents the ESG values for this set of stocks.

For an individual stock, the FY ESG values were smoothly interpolated
to provide daily values for $Z_{t_{n,k}}^{(\tX)}$.
As the FY ESG values for year Y were typically released at the beginning
of year $\textrm{Y}+1$, each FY ESG value was assigned to the last day of
the FY.
Daily values were computed in two steps: piecewise linear interpolation
followed by the application of a Gaussian--weighted, moving average
smoothing operation.
The result was ESG values $Z_{t_{n,k}}^{(\tX)}$ covering the time
period 31 December 2015 through 31 December 2023.
The additional ESG value needed for 02 January 2024 was obtained by
setting it equal to that for 31 December 2023.
See \citet[][Fig. 3]{asare-nyarko2025} for an illustration of this
two-step smoothing operation.

Computation of the ESG score $Z_{t_{n,k}}^{(\tM)}$ for the market index
requires knowledge of the company weights used in evaluating the index
value at time $t$.
Depending on the index, these weights (and indeed the index composition)
change somewhat regularly.
We were unable to obtain historical values for the component weights of
the Nasdaq-100,
but were able to obtain component weights for the Nasdaq-100--tracking ETF
Invesco QQQ Trust (QQQ) for the date 01 March 2024.\footnote{
	Source: www.slickcharts.com. Accessed 01 March 2024.
}
We therefore used the ETF QQQ as a proxy for the Nasdaq-100 index.
We computed daily ESG values $Z_{t_{n,k}}^{(\tM)}$ for QQQ by
applying the 01 March 2024 component weights to the daily ESG values
$Z_{t_{n,k}}^{(\tX)}$ of its component stocks and summing.

\subsection{Estimation of CSYIP values from QQQ}\label{sec:CSYIP}
Price values for QQQ were obtained for the period 04 January 2016 through
02 January 2024.\footnote{
	Source: Bloomberg Professional Services. Accessed 01 March 2024.
}
Given computed values of $Z_{t_{n,j}}^{(\tM)}$ (CSYIP property (i))
based on a set of $N$ consecutive historical trading days,
following \cite{Hu_2020} we estimate the probability $\mathbb{p}_{n,\Delta}^{(\tM)}$
by
$$
	\mathbb{p}_{n,\Delta}^{(\tM)}
		= \frac{\sum_{j=1}^N \text{I}_{ \left\{Z_{t_{n,k}}^{(\tM)} \geq 0 \right\} } }{N}.
$$
For our QQQ historical data set, we obtained the value
$\mathbb{p}_{n,\Delta}^{(\tM)} = 0.524$, yielding the binary values for the Bernoulli
random variable $\xi_{n,k}^{(\tM)}$,
\begin{equation*}
	\sqrt{ \frac{ 1-\mathbb{p}_{n,\Delta}^{(\tM)} }{ \mathbb{p}_{n,\Delta}^{(\tM)} } } = 0.954,
	\qquad
	\sqrt{ \frac{ \mathbb{p}_{n,\Delta}^{(\tM)} }{ 1-\mathbb{p}_{n,\Delta}^{(\tM)} } } = 1.05.
\end{equation*}

\begin{figure}[H]
\centering
	\includegraphics[width=0.8\textwidth]{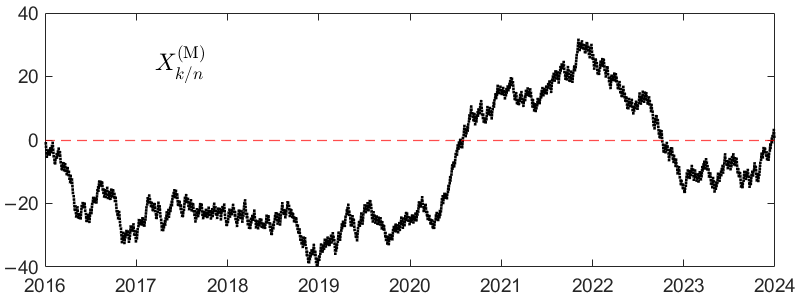}
	\caption{ The time series $X_{k/n}^{(\tM)}$ computed from the proxy index
		values $c_{t_{n,k}}^{(\tM)}$.
	}
	\label{fig:Xk}
\end{figure}
Fig.~\ref{fig:Xk} presents the time series $X_{k/n}^{(\tM)}$ (CSYIP property (iii)).
A critical choice in the application of this model is the form of the function
$h(x)$, $x \in \mathbb{R}$ used in the CSYIP property (iv).
\cite{Hu_2020b} explored the one-parameter form
$h(x) = (\sqrt{2 \pi} \sigma_h)^{-1} \exp(-x^2 / (2 \sigma_h^2)$.
They noted that this Gaussian form acts similar to a band--pass filter,
``letting through'' values of the argument $x$ that are within $2 \sigma_h$
of the value zero and
suppressing the values of the argument $x$ that are larger than 
$2 \sigma_h$ in magnitude.
We argue that this choice is incorrect for the following reason.
Fig.~\ref{fig:Xk} shows that the argument $X_{k/n}^{(\tM)}$ of $h(\cdot)$
took values within neighborhoods of zero for only four dates.
Furthermore, substantial change of $X_{k/n}^{(\tM)}$ ``away from zero''
occurred fairly rapidly with time.
Thus the Gaussian filter would effectively limit input to four short time
intervals.\footnote{
	The extent of this limitation depends on the value of the parameter
	$\sigma_h$.
}

\begin{figure}[H]
\centering
	\includegraphics[width=0.8\textwidth]{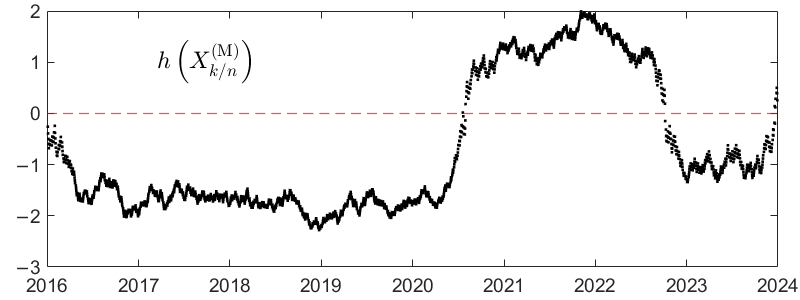}
	\includegraphics[width=0.8\textwidth]{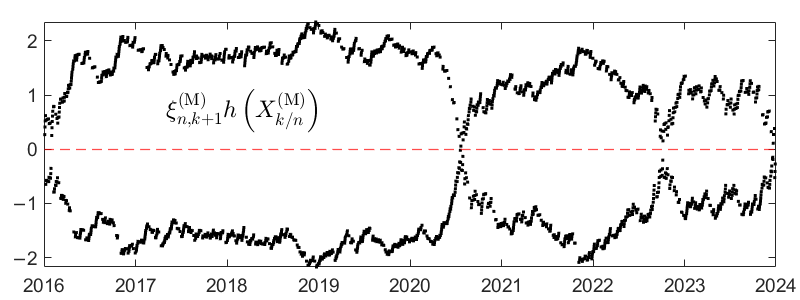}
	\caption{ The time series $h\left( X_{k/n}^{(\tM)} \right)$ and
		$\xi_{n,k+1}^{(\tM)} h\left( X_{k/n}^{(\tM)} \right)$
		computed from the proxy index QQQ.
	}
	\label{fig:Hk}
\end{figure}

We note that a value of $X_{k/n}^{(\tM)} = C > 0$ indicates that the market
index has had daily positive changes for at least $C$ days.
Thus, a large value of $X_{k/n}^{(\tM)} \gg 0$ is indicative of a strong market
history which will surely be reflected in stock prices.
Alternatively, a value of $X_{k/n}^{(\tM)} = C \ll 0$ indicates that the market
index has had daily decreases for at least $C$ days and
stock prices should reflect a decrease.
Thus, we reason that it is precisely the large values of $|x|$ that should not
be suppressed by $h(x)$.
Therefore, we have chosen the one-parameter form
\begin{equation}\label{eq:h_func}
	h(x) = \left( \frac{x}{d} \right)^{3/5}, \quad x_{\mathrm{min}} \leq x \leq x_{\mathrm{max}}.
\end{equation}
This choice has the property that $h(x)$ is nonlinear and does not grow
``too rapidly`` for small or large values of $|x|$.
Based on the maximum value of $ \left| X_{k/n}^{(\tM)} \right|$
for the proxy index QQQ, we settled on the parameter value $d = 10$;
resulting in values of
$\left| h\left( X_{k/n}^{(\tM)} \right) \right| \approx 1$.
Fig.~\ref{fig:Hk} shows the resultant time series for
$h\left( X_{k/n}^{(\tM)} \right)$ and
$\xi_{n,k+1}^{(\tM)} h\left( X_{k/n}^{(\tM)} \right)$ computed from the
$X_{k/n}^{(\tM)}$ time series in Fig.~\ref{fig:Xk}.

\subsection{Estimation of Regression Parameters}\label{sec:params}
We fitted the path--dependent, BBSM model \eqref{eq:c_tilde} to each
stock’s adjusted price series and estimated the model parameters
under three levels of ESG affinity, $\gamma^{\tESG} \in \{-1,0,1$\}.
We used the entire 2,012 days of price changes available
for each stock for the fits, obtaining constant parameter values
$a$, $\mu$, $v$, $\sigma$, and $\gamma$.
Fits were performed using a standard, constrained, linear least squares
optimization to minimize the $L_2$--norm of the residual
\begin{equation}\label{eq:lsq}
	\epsilon_{t_{n,k}} = c_{t_{n,k}} - \tilde{a}
		- \tilde{\mu} A_{t_{n,k}}
		- \tilde{v} \xi_{n,k+1}^{(\tM)}
		- \tilde{\sigma} A_{t_{n,k}} \xi_{n,k+1}^{(\tM)}
		- \tilde{\gamma} h\left( X_{k/n}^{(\tM)} \right) \xi_{n,k+1}^{(\tM)},
\end{equation}
subject to the constraint
\begin{equation}\label{eq:lsq}
	\tilde{v} + \tilde{\sigma} A_{t_{n,k}}
		+ \tilde{\gamma} h\left( X_{k/n}^{(\tM)} \right) \geq 0.
\end{equation}

Parameter and adjusted $R^2$ values for the fits with $\gamma^{\mathrm{ESG}} = 0$
are presented in Table~\ref{tab:params_app}.
For all 10 stocks, $\tilde{\sigma} A_{t_{n,k}} \xi_{n,k+1}^{(\tM)}$ is
the dominant term in the regression.
For eight stocks, the dominance ordering of the remaining terms is
$\tilde{v} \xi_{n,k+1}^{(\tM)}$,
$\tilde{\gamma} h\left( X_{k/n}^{(\tM)} \right) \xi_{n,k+1}^{(\tM)}$,
and $\tilde{a} + \tilde{\mu} A_{t_{n,k}}$.
For INTC and GOOGL, the ordering of
$\tilde{v} \xi_{n,k+1}^{(\tM)}$ and
$\tilde{\gamma} h\left( X_{k/n}^{(\tM)} \right) \xi_{n,k+1}^{(\tM)}$
is reversed.

\begin{figure}[H]
\centering
	\includegraphics[width=0.24\textwidth]{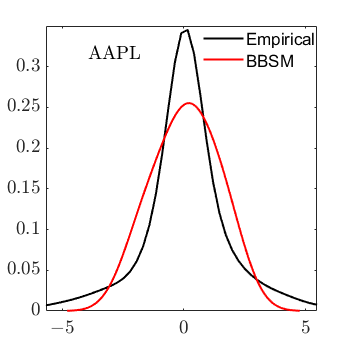}
	\includegraphics[width=0.24\textwidth]{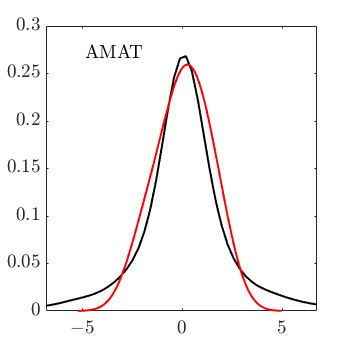}
	\includegraphics[width=0.24\textwidth]{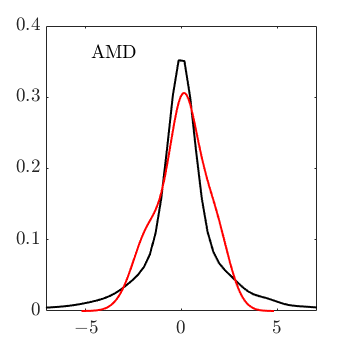}
	\includegraphics[width=0.24\textwidth]{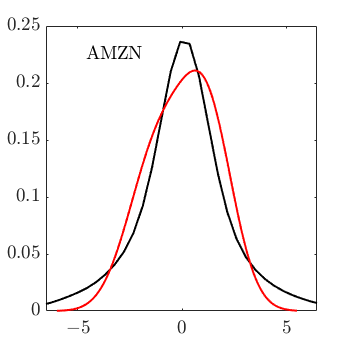}

	\includegraphics[width=0.24\textwidth]{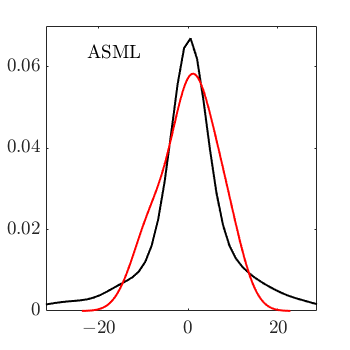}
	\includegraphics[width=0.24\textwidth]{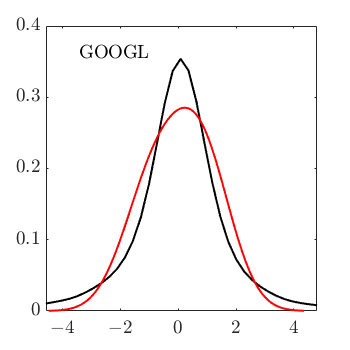}
	\includegraphics[width=0.24\textwidth]{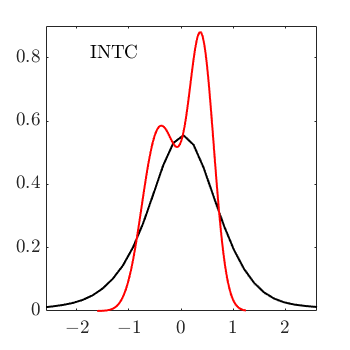}
	\includegraphics[width=0.24\textwidth]{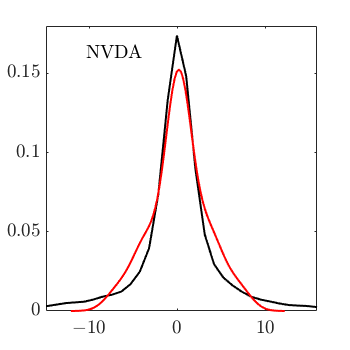}

	\includegraphics[width=0.24\textwidth]{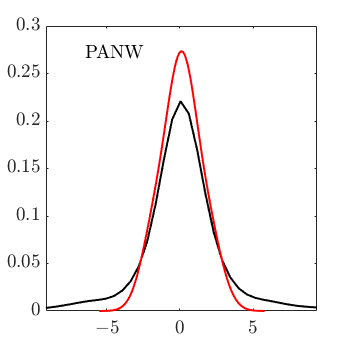}
	\includegraphics[width=0.24\textwidth]{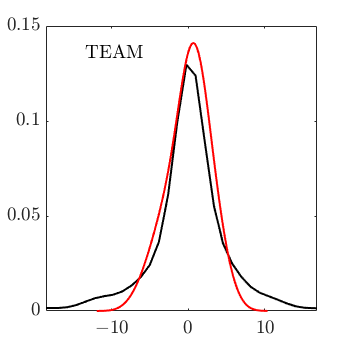}

	\caption{Estimated density of daily price changes for
		$\gamma^{\tESG} = 0,$.
	}
	\label{fig:pdf_0}
\end{figure}

Adjusted $R^2$ values vary from 17\% to 38\%; indicating relatively poor fits.
The reason can be traced to the Bernoulli variable $\xi_{n,k}^{(\tM)}$.
At a fixed value of $t_{n,k}$, an ensemble of discrete paths $X_{k/n}^{(\tM,j)}$,
$ j = 1, ..., m$, slowly converges--in--law to Brownian motion as $\Delta \downarrow 0$.
However, the rate of convergence is unknown.
Certainly, a single path $X_{k/n}^{(\tM)}$ determined from a single historical
set of daily market index data is not expected to be a good approximation to
Brownian motion.
For each stock, a kernel density fit to the empirical price change $c_{t_{n,k+1}}$
is compared to the kernel density of the regression fit in Fig.~\ref{fig:pdf_0}.
The density fits have been smoothed sufficiently to account for the bimodal
nature of the Bernoulli variable $\xi_{n,k+1}^{(\tM)}$.
The model cannot match the heavy--tailed nature of the empirical price change
distributions.

The parameters for the riskless asset price dynamics \eqref{eq:B_dyn}
were fit using linear least squares optimization, minimizing the
residual
\begin{equation}\label{eq:B_fit}
	\delta_{t_{n,k+1}} = \frac{\beta_{t_{n,k+1}} - \beta_{t_{n,k}}}{\Delta}
		- \rho - r \beta_{t_{n,k}}.
\end{equation}
The sequence of daily values $\beta_{t_{n,k}}$ required in \eqref{eq:B_fit}
was generated as follows,
\begin{equation}\label{eq:B_gen}
	\beta_{t_{n,k+1}} = (1 + r_{3,k} \Delta)\beta_{t_{n,k}},
	\qquad \beta_{t_{n,0}} = \beta_0,
\end{equation}
where $r_{3,k}$ is the three--month U.S. Treasury bill rate, converted to
a daily rate.\footnote{
	Data obtained for the period 04 January 2016 through 02 January 2024.
	Source: U.S. Department of the Treasury Daily Treasury Par Yield Curve
	Rates. Accessed 01 March 2024.
}
In essence, \eqref{eq:B_fit} models the evolution of the yield of the
three--month Treasury bill using the discrete form of the BBSM model
riskless rate dynamics \eqref{eq1.4}.

A critical question is the choice of the value of $\beta_0$.
In addressing this question we note the following.
Under time--independent constant coefficients, the solution \eqref{eq1.6}
reduces to
\begin{equation}\label{eq:cc_Bt}
	\beta_t = \left( \beta_0 + \frac{\rho}{r} \right) e^{rt} - \frac{\rho}{r}\,.
\end{equation}
and the solution to \eqref{eq:B_dyn} is
\begin{equation}\label{eq:Bk_soln}
	\beta_{t_{n,k}} = \left[ \beta_0 + \frac{\rho}{r}\right] \left( 1 + r \Delta\right)^k
			- \frac{\rho}{r}\,.
\end{equation}
We consider the question of how $\beta_t$  and $\beta_{t_{n,k}}$ scale with $\beta_0$.
Multiplying \eqref{eq:cc_Bt} by the constant $c$ gives
\begin{equation}\label{eq:c_B0}
	(c \beta_t )= \left( (c \beta_0)  + \frac{(c \rho)}{r} \right) e^{rt}
				- \frac{(c\rho)}{r}\,.
\end{equation}
Thus, if $\beta_0$ scales to $c \beta_0$, then $\beta_t$ scales to $c \beta_t$ and
$\rho$ scales to $c \rho$.
Note that $r$ does not change as  $e^{rt}$ does not scale.
Applying this analysis to the discrete equation \eqref{eq:Bk_soln} results
in the same conclusions.

Thus the value of the parameter $\rho$ obtained from the fit \eqref{eq:B_fit}
will depend on the value of $\beta_0$.
We  adjusted for this dependence by the reparameterization $\rho \to \rho \beta_0$.
Under the reparameterized form of \eqref{eq:B_fit},
\begin{equation}\label{eq:B_refit}
	\delta_{t_{n,k+1}} = \frac{\beta_{t_{n,k+1}} - \beta_{t_{n,k}}}{\Delta}
		- \rho \beta_0 - r \beta_{t_{n,k}}\,,
\end{equation}
both $\rho$ and $r$ are independent of the choice of $\beta_0$.
As values for the stock prices $A_{t_{n,k}}$ varied by more than one order
of magnitude,
when computing option prices we controlled for this variation
(e.g., controlling the magnitude of the ratio $A_{t_{n,k}} / \beta_{t_{n,k}}$
in \eqref{eq5.3})
by setting $\beta_0 = A_0$ and recomputing the sequence of values
$\beta_{t_{n,k}}$ for each stock and choice of $\gamma^{\tESG}$.

The parameter and adjusted $R^2$ values for the fit \eqref{eq:B_refit} are
$\rho = -0.00139$, $r = 0.00140$ with an adjusted $R^2$ value of 0.418.
Recall that the MB and BSM models exist in disjoint 3D parameter subspaces
of the BBSM model.
The parameter values listed above and in Table~\ref{tab:params_app} do not
lie in (or close to) either 3D subspace,
rejecting either the MB or BSM sub--models as appropriate for any of the
stocks.

Applying the same initial--value dependence analysis to \eqref{eq1.5}
shows that the parameters $a_t$ and $v_t$ scale linearly with $A_0$.
This holds for the parameters $a_{t_{n,k}}$, $v_{t_{n,k}}$ and
$\gamma_{t_{n,k}}$ in the discrete, path--dependent, BBSM model
\eqref{eq:c_tilde}.
Thus we adopt the reparameterizations
$a_{t_{n,k}} \to a_{t_{n,k}} A_0$,
$v_{t_{n,k}} \to v_{t_{n,k}} A_0$, and
$\gamma_{t_{n,k}} \to \gamma_{t_{n,k}} A_0$,
The values of the time--independent, $A_0$--independent parameters
(denoted $a / A_0$, $v / A_0$ and $\gamma / A_0$) computed from
Table~\ref{tab:params_app} are given in Table~\ref{tab:reparams}.
Unlike $r$ and the reparameterized value $\rho$ for the riskless asset,
reparameterized values (as well as $\mu$ and $\sigma$, which should not scale
with $A_0$) for the risky assets do exhibit slight variation with choice
of asset.
This may be a by-product of fitting for five parameters at once.

\subsection{Path--Dependent Option Price Valuation}\label{sec:op_prc}
Using the parameters estimated in Section~\ref{sec:params},
we computed European call option prices under the BBSM binary pricing tree
model \eqref{eq:BBSM_bin}, \eqref{eq:B_dyn}, \eqref{eq5.5}, \eqref{eq5.3}.
For each of the 10 selected stocks we computed option prices based upon the
underlying stock price on 03 January 2024.
Computed option prices were compared against weekly and monthly call option
data\footnote{
	Source: Yahoo Finance (finance.yahoo.com). Accessed on 03 January 2024.
}
published for 03 January 2024.
Option prices were computed for three values of the ESG affinity parameter,
$\gamma^{\tESG} \in \{-1,0,1\}$.
This is not meant to be a definitive investigation of the affect of the affinity
parameter, but rather to provide an indication of the effect of ESG rating on
the pricing of options under the BBSM model.

As the BBSM pricing tree is binary, the computational cost and memory
requirements for a given pair
of values $K,T$ is $\mathcal{O}(\sum_{t=0}^T 2^t)$, which greatly exceeds
the computational cost/memory requirements of $\mathcal{O}(\sum_{t=0}^T (t+1))$
for a binomial tree.
Using computational nodes with 512 GBytes of memory divided into groups of cores
each sharing 128 GBytes of memory,
with $\Delta = 1$ day we were able to compute to maturity times up to $T = 26$
days.
In order to achieve larger maturity times, we would have to increase the size of
$\Delta$, with commensurate modifications of $\mathbb{p}_{n,\Delta}^{(\tM)}$,
$\xi_{n,k}^{(\tM)}$, etc.

For the date 03 January 2024, published weekly and monthly option prices were
available for
$T \in \{2,7,11,17,21,26,31,49,73,116,135,179,260\}$ days for almost all stocks.
For a given value of $T$, the number of strike prices $K$ for which a call option
price was published varied with stock.
For example, for $T = 26$ days, the number of published strike/option prices
varied from two (ASML) to 36 (NVDA).
We therefore computed theoretical call option prices for $T \in \{2,7,11,17,21,26\}$
using $\Delta = 1$.
For each stock, the strike prices chosen for the theoretical computation were
determined from the full set of published contracts (over all maturities).

\begin{figure}[H]
	\centering
	\includegraphics[width=0.3\textwidth]{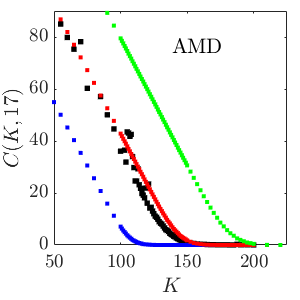}
	\includegraphics[width=0.3\textwidth]{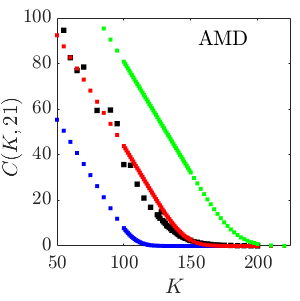}
	\includegraphics[width=0.3\textwidth]{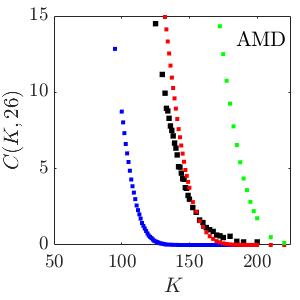}

	\includegraphics[width=0.3\textwidth]{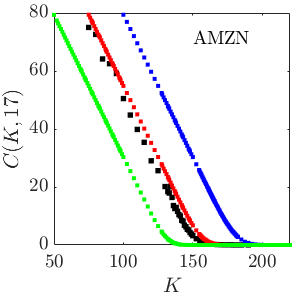}
	\includegraphics[width=0.3\textwidth]{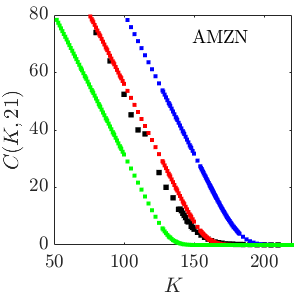}
	\includegraphics[width=0.3\textwidth]{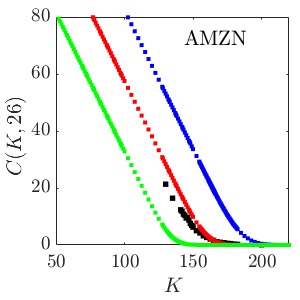}

	\caption{Empirical (black points) and theoretical (colored points) option
		prices $C(K,T)$ for $T \in \{17, 21, 26\}$ days and
$\gamma^{\mathrm{ESG}} \in \{-1 (\mathrm{blue}), 0 (\mathrm{red}),1 (\mathrm{green})\}$.
	}
	\label{fig:OP_txt}
\end{figure}

Fig.~\ref{fig:OP_txt} compares the empirical and theoretical option prices
$C(K,T)$ for $T \in \{17, 21, 26\}$ days and $\gamma^{\mathrm{ESG}} \in \{-1, 0,1\}$
for the stocks AMD and AMZM.
These two stocks provided reasonably large amounts of published strike/option
prices for these three values of $T$.
We concentrate first on the comparison between the empirical values of $C(K,T)$
and the $\gamma^{\mathrm{ESG}} = 0$ computations, which contain no ESG effect.
The fits at the three maturity times are quite good, although the BBSM model
does tend to overestimate the option price for in--the--money values of $K$.
The comparable plots to Fig.~\ref{fig:OP_txt} for the other eight stocks are
displayed in Appendix~\ref{app:OP_plts}, Fig.~\ref{fig:OP_app}.
The theoretical option prices match empirical values much better for
AMAT, ASML, GOOGL NVDA, and PANW (than for AMD and AMZN), although there is
a paucity of empirical prices at $T = 26$ for AMAT and ASML.

The inclusion of ESG ratings in the option valuation
with $\left | \gamma^{\mathrm{ESG}} \right | = 1$ has a significant effect
on the option price.
For fixed $K,T$ pair of values, $\gamma^{\mathrm{ESG}} = -1$ significantly
lowers the option price for AMD, while $\gamma^{\mathrm{ESG}} = 1$ significantly
raises the price.
For a fixed value of $T$, let $K_T(C,\gamma^{\mathrm{ESG}})$ denote the strike
price determined by a given value of option price $C$ and $\gamma^{\mathrm{ESG}}$.
We have, to reasonably good precision, that
\begin{equation}\label{eq:del_KT}
	\left| K_T(C,-1) - K_T(C,0) \right| \approx \left| K_T(C,1) - K_T(C,0)\right|.
\end{equation}
The effect is opposite for AMZN;
$\gamma^{\mathrm{ESG}} = -1$ raises the option price while
$\gamma^{\mathrm{ESG}} = 1$ lowers it.
Again we see that \eqref{eq:del_KT} holds.

The reason for the difference in option price changes with $\gamma^{\mathrm{ESG}}$
between these two stocks is simply related to the sign of $Z_t^{(\tX;\tM)}$ in
\eqref{eq:ESG_prc}.
For AMD, $Z_t^{(\tX)} > Z_t^{(\tM)}$ and $Z_t^{(\tX;\tM)}$ is positive;
for AMZN, $Z_t^{(\tX)} < Z_t^{(\tM)}$ and $Z_t^{(\tX;\tM)}$ is negative.
Thus ESG-valued prices for AMZN decrease when $\gamma^{\mathrm{ESG}} > 0$ increases
in magnitude.
For the remaining eight stocks, we see from Fig.~\ref{fig:OP_app} that
\eqref{eq:del_KT} holds.
The size of $\left| K_T(C,1) - K_T(C,0) \right|$ is stock dependent,
reflecting the stock dependence of $Z_t^{(\tX;\tM)}$.
GOOGL and TEAM had $Z_t^{(\tX;\tM)} < 0$, while the other six stocks had
$Z_t^{(\tX;\tM)} > 0$.

\section{Conclusion}\label{sec:conc}
We have integrated the history of the dynamics of a market index--following ETF
into a BBSM, dynamic, binary tree, pricing model.
Fitting the model parameters to eight years of historical data, we have demonstrated
that quite accurate fits to empirical option price data can be achieved from this
model, up to maturity times of 26 days.
Further scaling of our computations are limited by the computational cost and memory
requirements of binary--tree based calculations.

It would be of interest to perform the same option pricing computations and comparison
to empirical option prices under a
``pure'' BBSM model, without the path--dependency term.
Such a comparison was not performed in the BBSM paper of \cite{asare-nyarko2025}.
Given the time constraints on this manuscript and the computational time required
to repeat the binary tree computations, we leave this for further study.

Our computations revealed that the inclusion of ESG ratings into the price valuation
of a stock can have significant effect on option prices, either downward (if the market
sees a negative ESG affinity), or upward (under a positive market view of ESG).
While such an ESG--adjusted valuation does not affect the financial price of an option,
it would affect the choice of stock option to use as a hedging instrument in an
ESG--conscious portfolio.

\appendix
\appendixpage

\section{The Cherny--Shiryaev--Yor Invariance Principle}\label{app:CSYIP} 

We review the framework and key results of the CSYIP.

Let $\mathbb{N} = 1, 2, ...$ denote the natural numbers.
Let $\xi_k$, $k \in \mathbb{N}$
be a sequence of mean $0$, variance $1$, iid random variables.
For any $n \in \mathbb{N}$, define the scaled sequence
$$
\xi_0^{(n)} = 0, \qquad \xi_k^{(n)} =  \xi_k / \sqrt{n}, \quad k \in \mathbb{N};
$$
the ``time'' values $k/n$,  $k \in \mathbb{N}$;\footnote{
	Note that there is no restriction that $k \le n$.
}
and the cumulative sequence values
$$
X_{k/n}^{(n)} = \sum_{i=1}^{k} \xi_i^{(n)}, \quad k \in \mathbb{N}.
$$

\cite{Cherny_2003} define a function $h: \mathbb{R} \to \mathbb{R}$ to be piecewise
continuous if there exists a collection of disjoint intervals $J_m$,\footnote{
	Each $J_m$ can be closed, open, semi--open, or a point.
}
$m \in \mathbb{N}$
such that:
\begin{enumerate}
\item $\bigcup_{m=1}^{\infty} J_m = \mathbb{R}$;
\item For every compact interval $J$, there exists an $M \in \mathbb{N}$
		such that $J \subseteq \bigcup_{m=1}^{M} J_m$;
\item On each $J_m$, $h: J_m \to \mathbb{R}$ is continuous and has finite limits at
	those endpoints of $J_m$ which do not belong to $J_m$.
\end{enumerate}
We refer to such functions as CSY piecewise continuous.
Given a CSY piecewise continuous function $h(\cdot)$,
define the cumulative random sequence,
\begin{equation*}\label{eq:Ykn}
	Y_{k/n}^{(n)} = \sum_{i=1}^{k} h\left( X_{(i-1)/n}^{(n)}\right) \xi_i^{(n)},
	\qquad k \in \mathbb{N}.
\end{equation*}

Let $\mathbb{B}_t^{(n)}$, $t \geq 0$, be a random process with
piecewise linear trajectories having vertices $(k/n,X_{k/n}^{(n)})$.
Define $\mathbb{C}_t^{(n)}$, $t \geq 0$ to be a random process with
piecewise linear trajectories having vertices $(k/n, Y_{k/n}^{(n)})$, $k \in \mathbb{N}$.
From \citet[][Theorem 2.1]{Cherny_2003}:

\hfill\begin{minipage}{\dimexpr\textwidth-1cm}
If $h: \mathbb{R} \to \mathbb{R}$ is a CSY piecewise continuous function,\\
then, as $n \to \infty$, 
the bivariate process $(\mathbb{B}_t^{(n)}, \mathbb{C}_t^{(n)})$, $t \geq 0$ 
converges in law to $(B_t, C_t)$, $t \geq 0$, where $B_t$, $t \geq 0$, 
is a standard Brownian motion and $C_t = \int_{0}^{t} h(B_s) \, dB_s$.
\end{minipage}

\noindent
We refer to this theorem as the CSYIP.

\renewcommand{\thetable}{B\arabic{table}}
\setcounter{table}{0}

\section{Data Tables}\label{app:data_tabs}
\begin{table}[H]
	\begin{center}
	\caption{Summaries of the 10 stocks used in the empirical study.}
	\label{tab:stocks}
	\begin{tabular}{llll}
		\toprule
		Ticker & Company Name & GICS Sector & Headquarters \\
		\midrule
		AAPL & Apple Inc. & information technology & Cupertino, CA \\
		AMAT & Applied Materials, Inc. & information technology & Santa Clara, CA \\
		AMD  & Advanced Micro Devices, Inc. &  information technology & Santa Clara, CA \\
		AMZN & Amazon.com, Inc. & consumer discretionary & Bellevue, WA \\
		ASML & ASML Holding NV &   information technology & Veldhoven,  NL\\
		GOOGL & Alphabet, Inc. & communication services & Mountain View, CA \\
		INTC & Intel Corp. &   information technology & Santa Clara, CA \\
		NVDA & Nvidia Corp. & information technology & Santa Clara, CA \\
		PANW & Palo Alto Networks, Inc. &  information technology & Santa Clara, CA \\
		TEAM & Atlassian Corp. &  information technology & Sydney, AU \\
		\bottomrule
	\end{tabular}
	\end{center}
\end{table}

\begin{table}[H]
	\begin{center}
	\caption{Fiscal year ESG values for the set of stocks.}
	\label{tab:ESG}
	\begin{tabular}{l cccc cccc}
	\toprule
	FY		& 2015	& 2016	& 2017	& 2018	& 2019	& 2020	& 2021	& 2022\\
	\midrule
	AAPL	& 3.77	& 4.31	& 4.31	& 4.45	& 5.18	& 5.16	& 5.75	& 5.75\\
	AMAT	& 4.22	& 4.47	& 5.76	& 6.80	& 6.7	& 6.85	& 6.21	& 6.53\\
	AMD		& 5.25	& 5.92	& 6.02	& 6.11	& 6.56	& 6.64	& 6.28	& 6.24\\
	AMZN	& 1.91	& 2.43	& 2.48	& 2.38	& 3.52	& 3.98	& 4.15	& 4.17\\
	ASML	& 4.73	& 4.80	& 5.53	& 5.73	& 6.39	& 6.43	& 7.29	& 7.12\\
	GOOGL	& 2.87	& 2.58	& 2.78	& 3.18	& 3.55	& 4.31	& 4.35	& 4.30\\
	INTC	& 6.16	& 6.27	& 5.87	& 6.05	& 5.80	& 6.18	& 6.00	& 5.99\\
	NVDA	& 5.17	& 5.38	& 5.50	& 6.34	& 6.29	& 6.61	& 6.56	& 6.59\\
	PANW	& 1.01	& 1.22	& 1.31	& 1.20	& 2.17	& 3.29	& 4.81	& 5.63\\
	TEAM	& 0.90	& 1.27	& 1.41	& 1.56	& 3.06	& 2.93	& 3.11	& 2.92\\
	\bottomrule
	\end{tabular}
	\end{center}
\end{table}

\begin{table}[H]
	\begin{center}
	\caption{Regression parameter fits for $\gamma^{\mathrm{ESG}} = 0$.}
	\label{tab:params_app}
	\begin{tabular}{l lllll c}
		\toprule
		Stock & $\quad a$ & $\quad \mu$	   & $\quad v$ & $\quad \sigma$	& $\quad \gamma$ 
				& adj. $R^2$\\
		\ 	  & \   	& $\times 10^{-3}$ & \   & $\times 10^{-2}$ & \  & \ \\
		\midrule
		AAPL & 0.0666	&\ \ 0.210	&\ \ 0.418	&\ \ 0.796	&\ \ 0.172	& 0.384\\
		AMAT & 0.117	& $-$0.641	&\ \ 0.287	&\ \ 1.15	&\ \ 0.162	& 0.294\\
		AMD  & 0.0852	& $-$0.448	&\ \ 0.0731	&\ \ 1.73	&\ \ 0.0766	& 0.313\\
		AMZN & 0.271	& $-$2.19	&\ \ 0.293	&\ \ 1.01	&\ \ 0.126	& 0.331\\
		ASML & 0.564	& $-$675	&\ \ 1.92	&\ \ 0.999	&\ \ 1.01	& 0.327\\
		GOOGL & 0.0849	& $-$0.434	& $-$0.0179	&\ \ 1.15	&\ \ 0.0401	& 0.358\\
		INTC & 0.257	& $-$6.34	& $-$0.191	&\ \ 1.47	&\ \ 0.000845&0.204\\
		NVDA & 0.136	&\ \ 0.755	&\ \ 0.746	&\ \ 1.49	&\ \ 0.409	& 0.357\\
		PANW & 0.0168	&\ \ 1.01	& $-$0.0616	&\ \ 1.08	&\ \ 0.0800	& 0.171\\
		TEAM & 0.380	& $-$2.03	&\ \ 0.716	&\ \ 1.13	&\ \ 0.389	& 0.192\\
		\bottomrule
	\end{tabular}
	\end{center}
\end{table}

\begin{table}[H]
	\begin{center}
	\caption{Reparameterized values for $\gamma^{\mathrm{ESG}} = 0$.}
	\label{tab:reparams}
	\begin{tabular}{l lll}
		\toprule
		Stock & $a/A_0$ & $v/A_0$ & $\gamma/A_0$ \\
		\ 	  & $\times 10^{-2}$ & $\times 10^{-2}$ &  $\times 10^{-2}$ \\
		\midrule
		AAPL & 0.278	& 1.74		& 0.719\\
		AMAT & 0.699	& 1.72		& 0.969\\
		AMD  & 3.07		& 2.64		& 2.77\\
		AMZN & 0.850	& 0.920		& 0.394\\
		ASML & 0.703	& 2.39		& 1.25\\
		GOOGL & 0.224	& -0.0471	& 0.105\\
		INTC & 0.946	& -0.702	& 0.00311\\
		NVDA & 1.72		& 9.45		& 5.17\\
		PANW & 0.0294	& -0.108	& 0.140\\
		TEAM & 1.39		& 2.61		& 1.42\\
		\bottomrule
	\end{tabular}
	\end{center}
\end{table}

\renewcommand{\thefigure}{C\arabic{figure}}
\setcounter{figure}{0}

\section{Option Price Plots}\label{app:OP_plts}
\begin{figure}[H]
\centering
  \begin{subfigure}{\textwidth}
	\includegraphics[width=0.3\textwidth]{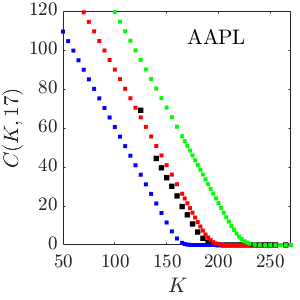}
	\includegraphics[width=0.3\textwidth]{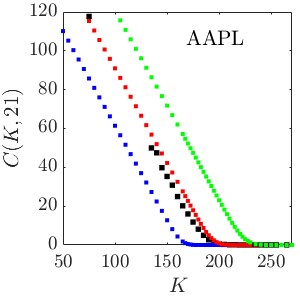}
	\includegraphics[width=0.3\textwidth]{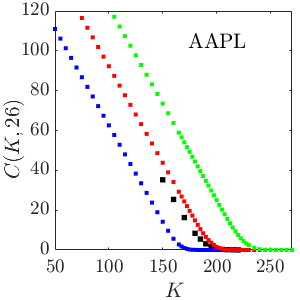}

	\includegraphics[width=0.3\textwidth]{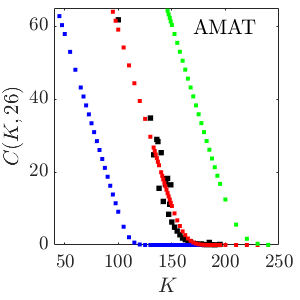}
	\includegraphics[width=0.3\textwidth]{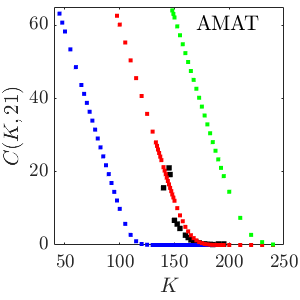}
	\includegraphics[width=0.3\textwidth]{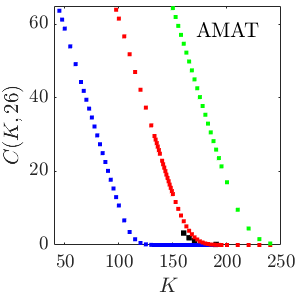}

	\includegraphics[width=0.3\textwidth]{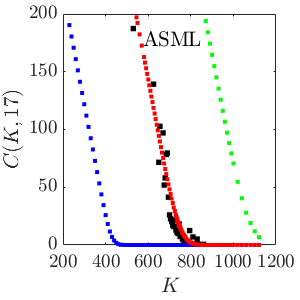}
	\includegraphics[width=0.3\textwidth]{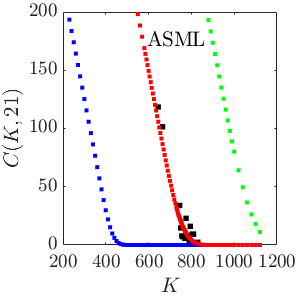}
	\includegraphics[width=0.3\textwidth]{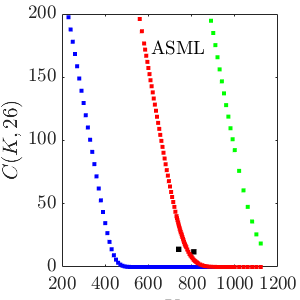}

	\includegraphics[width=0.3\textwidth]{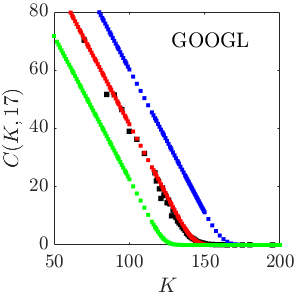}
	\includegraphics[width=0.3\textwidth]{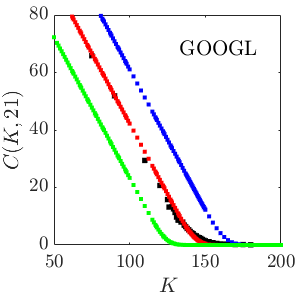}
	\includegraphics[width=0.3\textwidth]{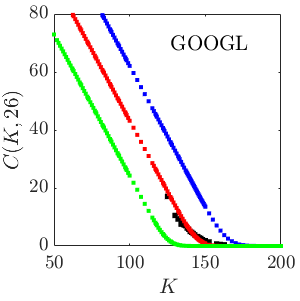}
  \end{subfigure}

	\caption{Empirical (black points) and theoretical (colored points) option
		prices $C(K,T)$ for $T \in \{17, 21, 26\}$ days and
	$\gamma^{\mathrm{ESG}} \in \{-1 (\mathrm{blue}), 0 (\mathrm{red}),1 (\mathrm{green})\}$.
	}
\end{figure}%
\begin{figure}[H]\ContinuedFloat
\centering
  \begin{subfigure}{\textwidth}
	\includegraphics[width=0.3\textwidth]{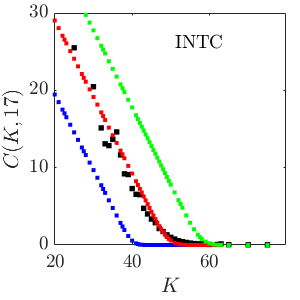}
	\includegraphics[width=0.3\textwidth]{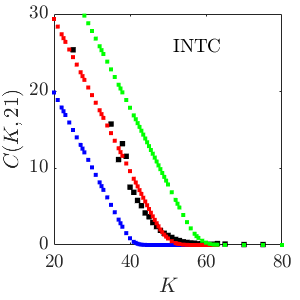}
	\includegraphics[width=0.3\textwidth]{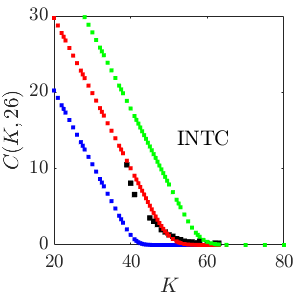}

	\includegraphics[width=0.3\textwidth]{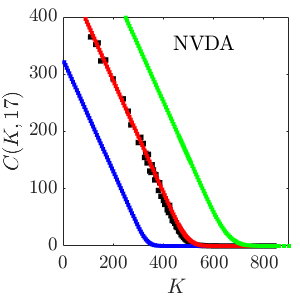}
	\includegraphics[width=0.3\textwidth]{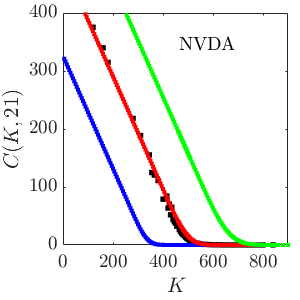}
	\includegraphics[width=0.3\textwidth]{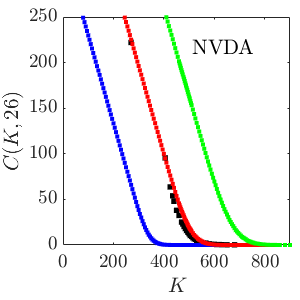}	

	\includegraphics[width=0.3\textwidth]{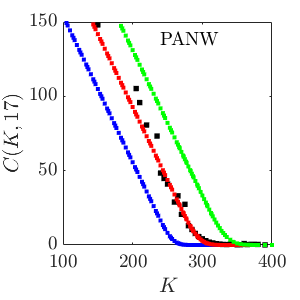}
	\includegraphics[width=0.3\textwidth]{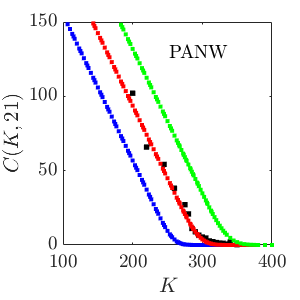}
	\includegraphics[width=0.3\textwidth]{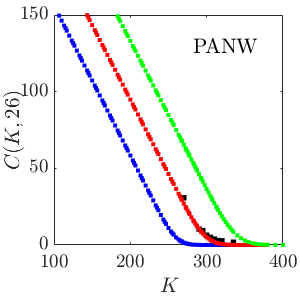}

	\includegraphics[width=0.3\textwidth]{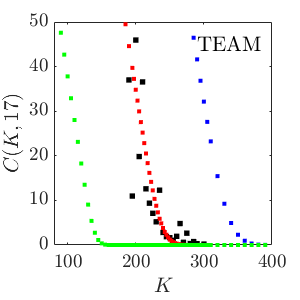}
	\includegraphics[width=0.3\textwidth]{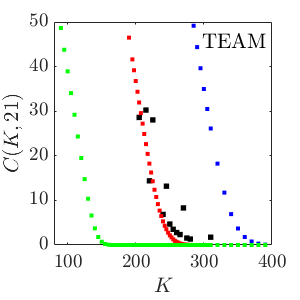}
	\includegraphics[width=0.3\textwidth]{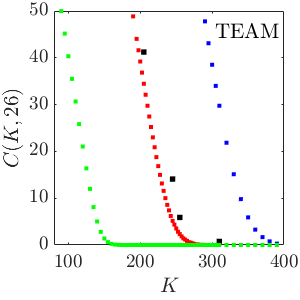}
  \end{subfigure}
	\caption{Empirical (black points) and theoretical (colored points) option
		prices $C(K,T)$ for $T \in \{17, 21, 26\}$ days and
	$\gamma^{\mathrm{ESG}} \in \{-1 (\mathrm{blue}), 0 (\mathrm{red}),1 (\mathrm{green})\}$.
	(cont.)
	}
	\label{fig:OP_app}
\end{figure}

\singlespacing
\normalem		
\bibliographystyle{chicago}
\bibliography{ESGpath}

\end{document}